\documentclass[12pt,a4paper]{article}
\usepackage{ifpdf}
\ifpdf
\pdfpageattr {/Group << /S /Transparency /I true /CS /DeviceRGB>>}
\fi
\usepackage{jheppub}
\usepackage{amsmath}
\usepackage{caption}
\usepackage{subcaption}
\usepackage{youngtab}

\newcommand{\beq}{\begin{equation}}
\newcommand{\eeq}{\end{equation}}
\newcommand{\bea}{\begin{eqnarray}}
\newcommand{\eea}{\end{eqnarray}}

\newcommand{\bra}{\langle}
\newcommand{\ket}{\rangle}

\newcommand{\eps}{\epsilon}

\begin{document}

\title{Low Tension Strings on a Cosmological Singularity}
\author[a,b]{Ben Craps,} \ 
\emailAdd{Ben.Craps@vub.ac.be}
\affiliation[a]{Theoretische Natuurkunde, Vrije Universiteit Brussel, and \\ International Solvay Institutes, \\ Pleinlaan 2, B-1050 Brussels, Belgium}
\affiliation[b]{Laboratoire de Physique Th\'eorique, \\ Ecole Normale Sup\'erieure, \\ 24 rue Lhomond, F-75231 Paris Cedex 05, France}
\author[c]{Chethan Krishnan,} \
\emailAdd{chethan.krishnan@gmail.com}
\affiliation[c]{Center for High Energy Physics, \\
  Indian Institute of Science, \\ 
  Bangalore - 560012, \ \ India}
\author[c,d,e]{Ayush Saurabh }
\affiliation[d]{International Center for Theoretical Sciences, \\
  Indian Institute of Science Campus, \\ 
  Bangalore - 560012, \ \ India}
\affiliation[e]{School of M. A. C. E., The University of Manchester,\\ 
  Manchester - M13 9PL, \ \ United Kingdom}
\emailAdd{ayushsaurabh@hotmail.com}
%\preprint{}  
\abstract{It has recently been argued that the singularity of the Milne orbifold can be resolved in higher spin theories. In string theory scattering amplitudes, however, the Milne singularity gives rise to ultraviolet divergences that signal uncontrolled backreaction. Since string theory in the low tension limit is expected to be a higher spin theory (although precise proposals only exist in special cases), we investigate what happens to these scattering amplitudes in the low tension limit. We point out that the known problematic ultraviolet divergences disappear in this limit. In addition we systematically identify all divergences of the simplest 2-to-2 tree-level string scattering amplitude on the Milne orbifold, and argue that the divergences that survive in the low tension limit have sensible infrared interpretations.
}

\setcounter{tocdepth}{2}
\maketitle

\section{Introduction}\label{introduction}

Singularities in time-dependent backgrounds are not well-understood in string theory. One context where some effort in this direction has been expended is in the case of time-dependent orbifolds of flat space; see \cite{Liu:2002yd, Cornalba:2003kd, Durin:2005ix, Craps:2006yb, Berkooz:2007nm} for reviews. Being descendants of flat space makes these geometries amenable to string perturbation theory, but it has been found that, already at tree level, string amplitudes in them have UV divergences arising from uncontrolled backreaction at the singularity \cite{LMS, Lawrence:2002aj, HoroPolch, Ben}. In this paper, we will revisit the string amplitudes in \cite{Ben} on the Milne geometry, which has a spacelike orbifold singularity.

Recently, it was observed in \cite{Shubho1, Shubho2} that certain cosmological singularities arising as quotients of dS$_3$ or flat space can in fact be resolved, if we embed these singularities into higher spin theories\footnote{The latter were formulated as Chern-Simons theories in 2+1 dimensions \cite{Rajesh, Campoleoni}.}. In particular, \cite{Shubho2} (see also \cite{Avinash}) shows that the Milne singularity can be resolved in the context of flat space higher spin theories \cite{Arjun, Troncoso}. The idea is that one can get rid of the metric singularity via higher spin gauge transformations, while preserving the holonomy in the Chern-Simons language.  

In the tensionless $\alpha' \rightarrow \infty$ limit of string theory, the tower of higher spin string states becomes massless. It is expected that this limit of string theory is captured by a massless, interacting higher spin theory: in AdS various arguments have been presented to make this correspondence more concrete \cite{Vasiliev, Sundborg, Witten,Giombi:2011kc,Chang:2012kt}. In the flat space case, it is unclear what the precise statement is, but one expects that the 2+1 D higher spin theory of \cite{Troncoso, Arjun} should morally capture some aspects of string theory in the tensionless limit, even in flat space.

If this belief is correct, one would expect that the string scattering amplitudes in Milne should be well-defined in the large $\alpha'$ limit. This is because the scattering amplitude is gauge invariant, so in the $\alpha' \rightarrow\infty$ limit it should be well-defined if the singularity is a gauge artifact in the higher spin picture. In this paper, we will do a scan of the divergences of the Milne 2-to-2 tree-level string scattering amplitude and show that it is indeed UV finite when the (dimensionless) $\alpha'$ is large enough. (We will be more precise about this in the next section.) Our analysis is exhaustive, and we will find infinite classes of divergences. However, we will argue that the divergences that survive in the large $\alpha'$ limit are all IR divergences that either (a) have been previously noticed in \cite{Ben}, or (b) can be understood to be arising from the tower of intermediate string states going on-shell and therefore are expected on physical grounds. 

The purpose of our analysis is to show that the problematic tree-level divergences identified in \cite{Ben} disappear in the large $\alpha'$ limit, which resolves the apparent tension between the pathological behavior of tree-level string scattering amplitudes on the Milne orbifold \cite{Ben} and the recent results suggesting that higher spin theory is well-behaved on the same space \cite{Shubho2}. The possible connection between string theory and higher spin theory suggests that string loop corrections should also be well-behaved in the  large $\alpha'$ limit, but this would be much harder to verify directly, and we will not attempt to do so here.

The paper is organized as follows. We first review the 2-to-2 tree-level string scattering amplitude on the Milne geometry (section 2) to fix our notations and to lay the groundwork for the discussions in the following sections. Section 3 describes various relevant features of the integrand and section 4 undertakes a careful scan of the divergences that can arise in this integral. This is the main technical part of the paper. Section 5 summarizes the various divergences and categorizes them as UV or IR. 

\section{Review: The 4-point String Amplitude on Milne}

\noindent
The Milne orbifold is obtained from Minkowski space $ds^2=-2dX^+ dX^-+d\vec X^2$ by the boost identification $X^\pm\rightarrow \exp(\pm 2\pi)X^\pm$. In \cite{Ben}, the four-point function $\bra \psi^*_3\psi^*_4\psi_1\psi_2 \ket$ of tachyon vertex operators
\bea\label{vertex}
\psi_{m_j,l_j,\vec p_j}= \frac{e^{i\vec p_j\cdot \vec X}}{2\sqrt2\pi i}\int_{-\infty}^{\infty}\,dw e^{\frac{i}{\sqrt2} (m_j X^-e^{-w} + m_j X^+e^w)}e^{i l_j w}
\eea
\noindent
 in tree-level bosonic string theory was computed to be
\bea\label{amplitude}
\sum \frac{(2\pi)^{24}}{4} \delta^{24} (\sum \eps_i \vec{p_i}) \delta (\sum \eps_i l_i) \int_{0}^{\infty} dv_4 G(s) G(t) G(u) \frac{v_2^{il_2+1} v_3^{-il_3+1} v_4^{-il_4-1}}{|m_2 m_3(v_2^2 - v_3^2)|}.
\eea
The  $\eps_i$ are $+1$ for the incoming particles (1 and 2), and $-1$ for the outgoing particles (3 and 4). The momenta $l_i$ along the Milne circle are integers and we will set them to zero in what follows because they are phases, and not crucial for the divergence/convergence discussion we undertake. Working with momenta measured in string units (which amounts to setting  $\alpha '=1$), the mass shell condition reads
\beq
m^2=-4 + \vec{p}^2.
\eeq
The parameter $m^2$ ($m>0$) is the effective 2-D mass squared. 
We have defined
\bea
G(x)=\frac{\Gamma (-1-\frac{x}{4})}{\Gamma (2+\frac{x}{4})}
\eea
and the Mandelstam variables are
\bea
s&=&-(p_1+p_2)^2=-8+m_1 m_2 (v_2+\frac{1}{v_2})-2\vec{p_1}.\vec{p_2},\\
t&=&-(p_1-p_3)^2=-8-m_1 m_3 (v_3+\frac{1}{v_3})+2\vec{p_1}.\vec{p_3},\\
u&=&-(p_1-p_4)^2=-8-m_1 m_4 (v_4+\frac{1}{v_4})+2\vec{p_1}.\vec{p_4}.
\eea
A standard constraint is
\bea
s+t+u=-16.
\eea
%in units where the dimensionful $\alpha'$ is set to 1. 
Also $v_2$ and $v_3$ are defined by\footnote{The upper and lower signs in the following expressions are correlated.}
\bea
v_2 & = &\frac{AB+m_2^2 - m_3^2 \mp \sqrt{\Delta}}{2 m_2 B}\label{sol1},\\
v_3 & = & \frac{-AB+m_2^2 - m_3^2 \mp \sqrt{\Delta}}{2 m_3 B}\label{sol2},
\eea
where
\bea
A &=& -m_1 + m_4 v_4 \,,\label{A} \\ B &=& -m_1 + \frac{m_4}{v_4} \,, \label{B}\\ \Delta &=& (m_2^2 - m_3^2)^2 - 2AB (m_2^2 + m_3^2) +A^2 B^2. \label{delta}
\eea
This specific form of $v_2$ and $v_3$ arises from delta functions that enforce 
\bea 
m_1+m_2 v_2-m_3 v_3-m_4 v_4&=&0,  \\  m_1+\frac{m_2}{v_2}-\frac{m_3}{v_3}-\frac{m_4}{v_4}&=&0.  
\eea
It is also important that $v_2,v_3$ need to be positive, so only the positive solutions need to be retained (and summed over) in the integral. The outermost summation in the amplitude (\ref{amplitude}) refers to this summation over the positive branches of $v_2$ and $v_3$.

The problematic divergence identified in \cite{Ben} comes from the large $v_4$ region of the integral (\ref{amplitude}), in which the integrand is in the Regge regime, $s\rightarrow m_1m_4v_4\rightarrow\infty$ with finite $t\rightarrow-(\vec p_1-\vec p_3)^2$. This integration region gives a contribution of the form
\bea
\int^\infty dv_4\, v_4^{-\frac{(\vec p_1-\vec p_3)^2}{2}},
\eea
which diverges whenever $\alpha'(\vec p_1-\vec p_3)^2 < 2$, where we reinstated $\alpha'$. This makes it clear that the problematic divergence disappears whenever the momentum transfer is large enough in string units, and therefore in the large $\alpha'$ limit. The physical intuition is that for large momentum transfer, string amplitudes are very soft in the Regge regime. 

Our goal in the rest of the paper is to study the convergence properties of (\ref{amplitude}) thoroughly, in order to make sure that no problematic divergences remain in the tree level amplitude. The integral has a fair amount of structure and the problem is fairly detail oriented, so in the next section we proceed systematically to characterize the integral.

As mentioned in the introduction, a study of string loop corrections is beyond the scope of the present paper. In Minkowski space, arguments have been presented that string loop diagrams are also soft in the Regge regime for sufficiently large momentum transfer \cite{Gross:1987kza} (although the perturbation series could not be summed). In the Milne orbifold, one would also have to include contributions from the exchange of twisted sector strings, which we will not attempt.

\section{Structure of the Integral}

 As stated earlier, we need pairs of positive $v_2$ and positive $v_3$ among (\ref{sol1}-\ref{sol2}), out of the four possible pairs of combinations. This means we have to find out the regions of integration in (\ref{amplitude}) 
where the positivity properties of $v_2$ and $v_3$ change. 
 
It turns out that there are three qualitatively different regions in the $v_4$-half line. The defining features of these regions are governed by the parameters $m_1$, $m_2$, $m_3$ and $m_4$. These parameters control where $v_2$ and $v_3$ change signs, or become complex, or are indeterminate. As we find out from (\ref{sol1}-\ref{sol2}), constituents of one of the pairs of $v_2$, $v_3$ change signs whenever $A=0$, and both $v_2$, $v_3$ become indeterminate whenever $B=0$. This defines the formal boundaries of the three adjoining regions of integration.

To understand these regions in more detail, we first introduce
\bea
v_{2u} & = &\frac{AB+m_2^2 - m_3^2 - \sqrt{\Delta}}{2 m_2 B}, \label{solu1} \\
v_{2d} & = &\frac{AB+m_2^2 - m_3^2 + \sqrt{\Delta}}{2 m_2 B}, \\
v_{3u} & = & \frac{-AB+m_2^2 - m_3^2 - \sqrt{\Delta}}{2 m_3 B}, \\
v_{3d} & = & \frac{-AB+m_2^2 - m_3^2 + \sqrt{\Delta}}{2 m_3 B}. \label{solu2}
\eea 
We also define some new parameters:
\bea 
P_{12} = \vec{p_1}.\vec{p_2}\,, \,\,P_{13} = \vec{p_1}.\vec{p_3}\,, \ \,P_{14} = \vec{p_1}.\vec{p_4},\\ 
v_{4A}=\frac{m_1}{m_4} \,,\,\, v_{4B}=\frac{m_4}{m_1}. \hspace{0.6in}
\eea
Now, it is easily inferred from (\ref{A}-\ref{B}) that $A$ changes sign at some $v_4=v_{4A}$ and $A>0$ for $v_4>v_{4A}$. Similarly, $B$ changes sign at some $v_4=v_{4B}$ and $B<0$ for $v_4>v_{4B}$. 

Depending on the actual values of $m_1$, $m_2$, $m_3$, $m_4$, $P_{12}$, and $P_{13}$, 
we will have specific kinematic conditions defining the regions of integration and the properties of the four-point function. These are what we call the kinematic parameters, and they fully describe the amplitude\footnote{Note that we have set the $l_i$ to zero as discussed earlier.}.

For all these cases, we adopt a global convention that the $m_1$ would always stand for the lighter of the two incoming particles (i.e. $m_2\ge m_1$) and the $m_3$ would always stand for the lighter of the two outgoing particles (i.e. $m_4\ge m_3$). Even with this assumption, we still have four separate cases of orderings betweeen the mass parameters that we can consider. Of these we call $m_4>m_1$ and $m_3>m_2$ Case-1, and $m_4>m_1$ and $m_2>m_3$  Case-2 and discuss them in detail. The other two cases (e.g., $m_1 > m_4$ and $m_2 > m_3$) are either kinematically impossible or have analogous divergence structures, so we can omit them.

It is also worth noticing that the parameters that we choose need to satisfy certain constraints due to the rules of vector algebra and momentum conservation:
\bea
\vec{p_1}^2+\vec{p_2}^2 \ge 2|P_{12}| &\Longleftrightarrow & |P_{12}| \le \frac{m_1^2 + m_2^2}{2} + 4, \label{firstcon} \\
\vec{p_1}^2+\vec{p_3}^2 \ge 2|P_{13}| &\Longleftrightarrow & |P_{13}| \le \frac{m_1^2 + m_3^2}{2} + 4 ,\\
\vec{p_1}^2+\vec{p_4}^2 \ge 2|P_{14}| &\Longleftrightarrow & |P_{14}| \le \frac{m_1^2 + m_4^2}{2} + 4, \\
\vec{p_1}+\vec{p_2} = \vec{p_3}+\vec{p_4} &\Longleftrightarrow & P_{14}=m_1^2+4-P_{13}+P_{12} . \label{lastcon}
\eea
\noindent
Note that $P_{14}$ is not an independent parameter.

Now we turn to a discussion of the two cases.

\subsection{Case-1: $m_4>m_1$ and $m_3>m_2$}

\noindent
In this case, $v_{4A}<v_{4B}$ and so we could define our three regions of integration as follows,\\ \\
Region-I\,\,\,\,\,: $v_4\in(0,v_{4A})$\\
Region-II\,\,\,: $v_4\in (v_{4A},v_{4B})$\\
Region-III\,: $v_4\in (v_{4B},\infty)$\\

\noindent
In Region-I, $A<0$ and $B>0$. So, through inspection of the expressions (\ref{solu1}-\ref{solu2}), we find that $\sqrt{\Delta}>(|AB|+|m_2^2 - m_3^2|)$. Therefore in Region-I, $v_{2u}$ and $v_{3u}$ would always be negative, and $v_{2d}$ and $v_{3d}$ would always be positive. Conversely, in Region-III, through similar analysis we find that $v_{2u}$ and $v_{3u}$ would always be positive, and $v_{2d}$ and $v_{3d}$ would always be negative. In Region-II, though, we find that we do not have any pair of $v_{2}$ and $v_{3}$ in (\ref{solu1}-\ref{solu2}) in which both constituents of the pair are positive at the same time. Hence, in the present case, the integrand does not exist in Region-II. 

For notational convenience, we will define the integrands in Region-I, III as follows:
\bea
i_1 \big(v_{2d}(v_4), v_{3d}(v_4), v_4)\big)= G(s(v_{2d})) G(t(v_{3d})) G(u(v_{4})) \, \frac{v_{2d} v_{3d} v_4^{-1}}{|m_2 m_3(v_{2d}^2 - v_{3d}^2)|}, \\
i_3 \big(v_{2u}(v_4), v_{3u}(v_4), v_4)\big)= G(s(v_{2u})) G(t(v_{3u})) G(u(v_{4})) \, \frac{v_{2u} v_{3u} v_4^{-1}}{|m_2 m_3(v_{2u}^2 - v_{3u}^2)|}.
\eea
\begin{figure}[h]
\centering
\includegraphics[width=0.8\textwidth]{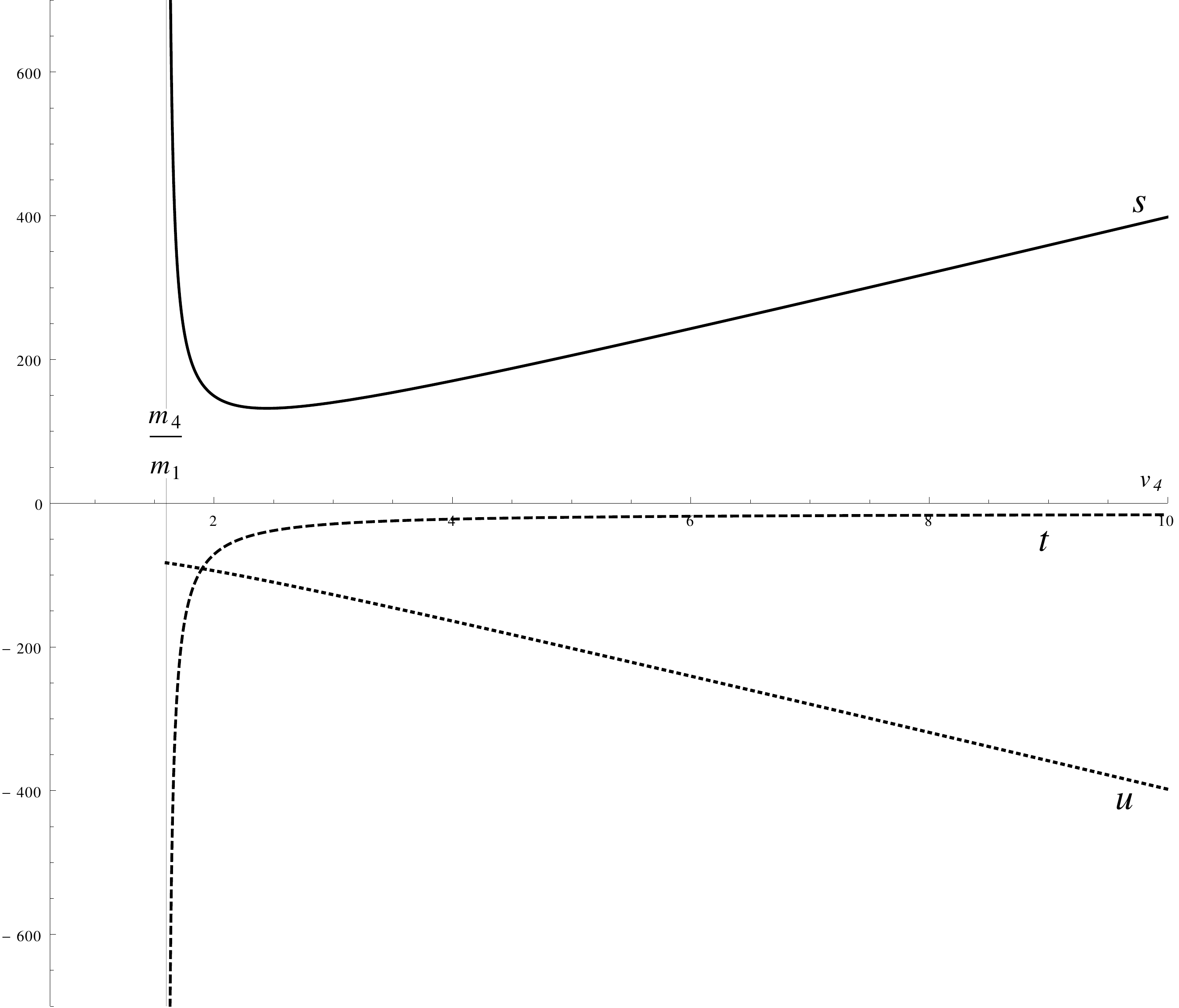}
\caption{Plots of $s$,$t$, and $u$ for $m_4 > m_1$, $m_3>m_2$ for positive $v_4$. The values of the kinematic parameters used for making the plot are $m_1=5.0,\ m_2=5.8,\ m_3=6.8,\ m_4=8.0,\ P_{12}=10.5, \ P_{13}=33.5$.}
%\label{fig:circles}
\end{figure}
\noindent
Now, the integral in Region-I (we call it $I_1$) can be easily shown to be the same as the integral in Region-III by an appropriate renaming of $v_4 \rightarrow \frac{1}{x}$. Specifically, the contribution to the four-point function integral from Region-III (we call it $I_3$) transforms as
\bea
& & I_3 = \sum \frac{(2\pi)^{24}}{4} \delta^{24} (\sum \eps_i \vec{p_i}) \delta (\sum \eps_i l_i) \int_{v_{4B}}^{\infty} dv_4 \, i_3 \big(v_{2u}(v_4), v_{3u}(v_4), v_4)\big) \nonumber \\
&\rightarrow& \,\, \sum \frac{(2\pi)^{24}}{4} \delta^{24} (\sum \eps_i \vec{p_i}) \delta (\sum \eps_i l_i) \int_{v_{4A}}^{0} dx \,\, i_3 \big(v_{2u}(x^{-1}), v_{3u}(x^{-1}), x^{-1})\big) \times -\bigg(\frac{1}{x^2}\bigg). \nonumber
\eea
\\
\noindent
Now if after simplifications we replace the $x$ with $v_4$, we find that the transformed integral above becomes
\bea
I_3 \,\rightarrow \, \sum \frac{(2\pi)^{24}}{4} \delta^{24} (\sum \eps_i \vec{p_i}) \delta (\sum \eps_i l_i) \int_{0}^{v_{4A}} dv_4 \,\, i_1 \big(v_{2d}(v_4), v_{3d}(v_4), v_4)\big) \,=\, I_1. \nonumber
\eea

\noindent
Thus the net scattering amplitude in this case is given by
\beq
I=I_1+ I_3 = 2 \,I_3. \nonumber
\eeq
In Figure-1 we have transformed away Region-I into Region-III and only show Region-III. 

\subsection{Case-2: $m_4>m_1$ and $m_2>m_3$}

In this case, we can choose the various regions as in Case-1. The integrand behaves very similar to the integrand in Case-1 in the $v_4 \rightarrow \infty$ limit but the difference now is that $s(v_{2u})$, $t(v_{3u})$, and $u$ are now finite in the $v_4 \rightarrow v_{4B}$ limit and the integrand now exists in Region-II as well. 

One crucial observation is that there are two possibilities to be distinguished. These two possibilities\footnote{Throughout the paper, we will be making comments about the case $(m_1+m_2) = (m_3+m_4)$, which can be thought of as a special case of either of the possibilities.} are $(m_3+m_4) \ge (m_1+m_2)$ and  $(m_1+m_2) \ge (m_3+m_4)$. If the former condition is satisfied, $\Delta$ can change sign in Region-II and values become complex. So $v_4$ corresponding to $\Delta = 0$ defines two boundaries (we call them $v_{4+}$ and $v_{4-}$)\footnote{There are four roots for the locations of $\Delta=0$ points in $v_4$ space. Out of these four roots, only two lie in $v_4>0$. We quote them here for completeness:
\beq
v_{4\pm}=\frac{m_1^2-m_2^2-m_3^2+m_4^2+2 m_2 m_3\pm\sqrt{\left(m_1^2-m_2^2-m_3^2+m_4^2+2 m_2 m_3\right){}^2-4 m_1^2 m_4^2}}{2 m_1 m_4}. \nonumber
\eeq}  
for the integral within Region-II as can be seen in the Figure-2.
\begin{figure}[h]
\centering
\includegraphics[width=0.8\textwidth]{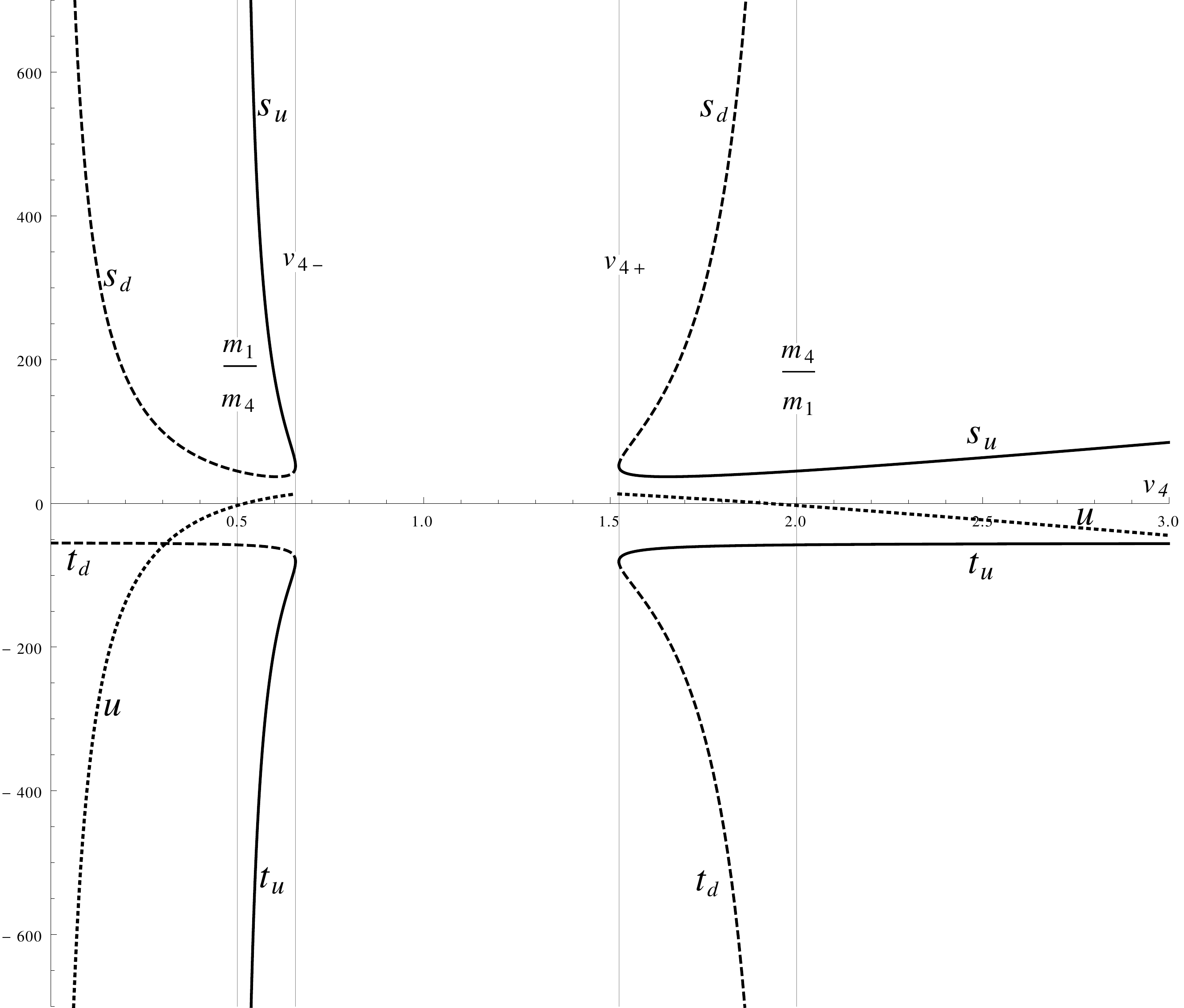}
\caption{Plots of $s$, $t$ and $u$ for $m_4 > m_1$, $m_2>m_3$, $(m_3+m_4) > (m_1+m_2)$ for  positive $v_4$, where the subscript $u$ stands for functions of $v_{2u}$, $v_{3u}$ and subscript $d$ stands for functions of $v_{2d}$, $v_{3d}$. The values of the kinematic parameters used for making the plot are $m_1=5.0,\ m_2=9.5,\ m_3=5.5,\ m_4=10.0,\ P_{12}=39.5, \ P_{13}=4.0$.}
%\label{fig:circles}
\end{figure}
%\noindent
We also find that all the solutions for $v_{2u}$, $v_{2d}$, $v_{3u}$, and $v_{3d}$ are either positive or complex, all at the same time for  any specific $v_4$-location in Region-II. Therefore, since $i_1(v_{2d},v_{3d}, v_4)$ and $i_3(v_{2u},v_{3u}, v_4)$ both exist at the same time in Region-II and could also be transformed into each other using the transformation process presented in Case-1, we would essentially have to worry only about integration in the interval $(v_{4+},\infty)$, which takes the form
\beq
I \, =\, 2\,\int_{v_{4+}}^{v_{4B}} dv_4\,  i_1 \big(v_{2d}(v_4),v_{3d}(v_4), v_4 \big) + 2\,\int_{v_{4+}}^{\infty} dv_4\,  i_3 \big(v_{2u}(v_4),v_{3u}(v_4), v_4 \big). \label{case2intsum}
\eeq

When $(m_1+m_2) > (m_3+m_4)$, however, $\Delta$ stays positive throughout Region-II. This case is shown in Figure-3.
\begin{figure}[h]
\centering
\includegraphics[width=0.8\textwidth]{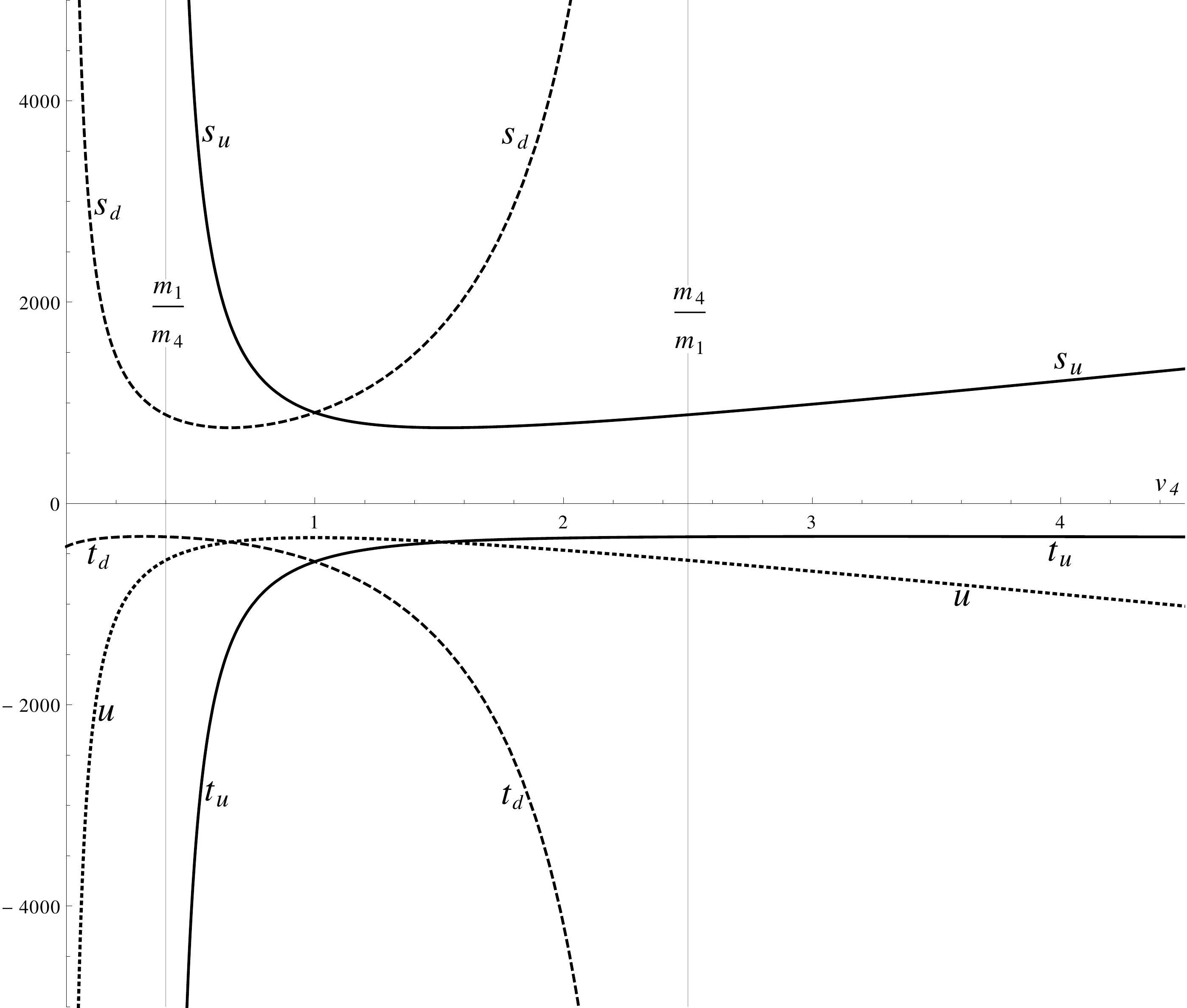}
\caption{Plots of $s$, $t$ and $u$ for $m_4 > m_1$, $m_2>m_3$, $(m_1+m_2) > (m_3+m_4)$ for  positive $v_4$, where the subscript $u$ stands for functions of $v_{2u}$, $v_{3u}$ and subscript $d$ stands for functions of $v_{2d}$, $v_{3d}$. The values of the kinematic parameters used for making the plot are $m_1=10.0,\ m_2=40.0,\ m_3=20.0,\ m_4=25.0,\ P_{12}=20.0, \ P_{13}=40.0$.}
%\label{fig:circles}
\end{figure}
%For the situation when $(m_4+m_3) \le (m_1+m_2)$, 
Now we always have $\Delta \ge 0$ and so the Mandelstam invariants are real in whole of Region-II. This means that the net integral simply becomes
\beq
I \, =\, 2\,\int_{1}^{v_{4B}}dv_4\,  i_1 \big(v_{2d}(v_4),v_{3d}(v_4), v_4 \big) + 2\,\int_{1}^{\infty}dv_4\,  i_3 \big(v_{2u}(v_4),v_{3u}(v_4), v_4 \big). \label{case2intsum2}
\eeq

\section{Divergences of the four-point function}

\noindent
In this section we will present a systematic scan of the divergences in the four-point function. This section is technical, but our conclusions are summarized in the next section for the reader's convenience. 

In the four-point function, we have multiple gamma functions which give rise to poles in the integrand. These poles are avoided by the $i\epsilon$ prescription of the Feynman propagator. This prescription differs from the Cauchy Principal Value (CPV) prescription by a delta function term, which does not give rise to a divergence. Therefore, since we are only interested in determining whether the integral is finite, it suffices to do so using a CPV prescription. 

For a generic gamma function, $\Gamma (x)$, around the pole $x=-n$ ($n>0$), we have
\beq
\lim\limits_{\delta x \rightarrow 0} \Big(\Gamma (-n-\delta x) + \Gamma (-n+\delta x) \Big) = \frac{2 (-1)^{-n} \psi (n+1) }{n!},
\eeq
where $\psi$ is the digamma function. 
The last expression is completely finite and would imply that the integration through a gamma function pole would be finite in the sense of CPV.

Now suppose the argument of the gamma function is a function $f(x)$, with $f(x)=-n$ for some $x=x_n$. If $f'(x_n)\neq 0$, a very similar argument shows that the integral around the pole is finite. However, if $x_n$ is a local minimum or a local maximum of $f$, then the pole is approached from the ``same side" and the integral diverges.

To use these observations in the analysis of our four-point function, we first write the four-point function integral around a pole (at say $v_4 = v_{4n}$) in the following form:
 \beq
 \int_{v_{4n}-\delta v_4}^{v_{4n}+\delta v_4} dv_4 \, F(v_4) \, \Gamma(M(v_4)) ,
 \eeq 
 
 \noindent
 where $M(v_4)$ is a function of one of the Mandelstam invariants, and $F(v_4)$ is assumed to be continuous 
in the concerned  interval, which would be the generic situation\footnote{If $F(v_4)$ passes through zero at $v_{4n}$, one might worry that the sign change would cause a divergence. But the pole of the gamma function goes as $1/(v_4 - v_{4n})$ and therefore any fractional or integer power law approach to zero of $F(v_4)$ (which are the only cases relevant for us), will result in a converging integral. A more tricky situation arises when $F(v_4)$, instead of being continuous, has a discontinuity at $v_{4n}$ and has opposite signs on either side of the pole. This can happen when more than one gamma function has a pole at the same location. But one can show that because of the $s+t+u=-16$ constraint, the integral is again finite: this is tied crucially to the Veneziano (or Virasoro-Shapiro) structure of the Gamma functions.}. By the above reasoning, poles at generic values of $v_4$ do not make the integral diverge.
However for some specific kinematic configurations we would have special gamma function poles which cause the four-point function to diverge. This can happen when 
\begin{itemize}
\item there is a pole at the boundary of integration so that there are no two ``sides" for the poles to cancel against each other in the CPV, or 
\item there is a pole at a maximum/minimum of the Mandelstam variable that falls within the integration range, as we discussed above.
\end{itemize}

Now, we classify and explain all the divergences of the four-point function as follows:
\\ \\
\noindent
Type-1: Divergences from boundaries of the integral not related to poles.\\
Type-2: Divergences from gamma function poles at the boundary of the integral. \\
Type-3: Divergences from gamma function poles occurring at the maxima/minima of Mandelstam invariants.\\
Type-4: IR type divergences in specific kinematic configurations.\\

\subsection{Type-1: Divergences from Boundaries, Unrelated to Poles}
For the kinematic configuration in Case-1, presented in the previous section, the two boundaries of the integral are $v_4=v_{4B}$ and $v_4=+ \infty$. In the $v_4 \rightarrow \infty$ limit, we find that 
\bea
& & v_{2u} \rightarrow \frac{(-m_1+m_4 v_4)}{m_2} \rightarrow \infty \,\,\, , \,\, v_{3u} \rightarrow \frac{m_3}{m_1}, \nonumber \\ \nonumber \\
& & s \rightarrow m_1 m_4 v_4 \rightarrow \infty \,\,\, , \,\, t \rightarrow -(\vec{p_1}-\vec{p_3})^2 \, , \,\,\, u \rightarrow -m_1 m_4 v_4 \rightarrow -\infty. \nonumber
\eea
This is the Regge limit condition, $s \rightarrow \infty$ and $t$ fixed. Under this condition the term $G(s)G(t)G(u)$ simplifies as
\beq
G(s)G(t)G(u) \rightarrow - \Big(\frac{s}{4}\Big)^{2+\frac{t}{2}} \, \frac{\Gamma(-1-\frac{t}{4})}{\Gamma(2+\frac{t}{4})}.\nonumber 
\eeq
\noindent
Using the limiting expressions above, the four-point function simplifies as 
\bea
 C \times \, \int_{}^{\infty} dv_4 \, v_4^{-\frac{(\vec{p_1}-\vec{p_3})^2}{2}} ,
\eea
\noindent
where $C$ is some constant. So in the $v_4 \rightarrow \infty$ limit, the four-point function will diverge whenever the exponent in the integrand is such that
\cite{Ben}
\beq
\frac{(\vec{p_1}-\vec{p_3})^2}{2} \le 1 \, \, \Longleftrightarrow  \,\, (\vec{p_1}-\vec{p_3})^2 \le 2.\nonumber
\eeq
\noindent
Similarly in $v_4 \rightarrow v_{4B}$ limit, we have
\bea
& & v_{2u} \rightarrow \frac{(m_2^2-m_3^2)v_4}{m_2(m_4-m_1v_4)} \rightarrow \infty \,\,\, , \,\, v_{3u} \rightarrow \frac{(m_2^2-m_3^2)v_4}{m_3(m_4-m_1v_4)} \rightarrow \infty, \nonumber \\ \nonumber \\
& & s \rightarrow \frac{m_1(m_2^2-m_3^2)v_4}{(m_4-m_1v_4)} \rightarrow \infty \,\,\, , \,\, t \rightarrow -\frac{m_1(m_2^2-m_3^2)v_4}{(m_4-m_1v_4)} \rightarrow \infty, \nonumber \\ \nonumber \\
& & u \rightarrow -(\vec{p_1}-\vec{p_4})^2. \nonumber  \nonumber 
\eea
\noindent
Again we have the Regge limit condition, $s \rightarrow \infty$ and $u$ fixed, which leads to the simplification
\beq
G(s)G(t)G(u) \rightarrow - \Big(\frac{s}{4}\Big)^{2+\frac{u}{2}} \, \frac{\Gamma(-1-\frac{u}{4})}{\Gamma(2+\frac{u}{4})}.\nonumber 
\eeq
\noindent
Using the limiting expressions above, the four-point function simplifies as 
\bea
 C^{'} \times \, \int_{v_{4B}}^{} dv_4 \, \bigg(\frac{1}{m_4-m_1 v_4} \bigg)^{2-\frac{(\vec{p_1}-\vec{p_4})^2}{2}} ,
\eea
\\
\noindent
where again $C^{'}$ is a constant. Thus, in $v_4 \rightarrow v_{4B}$ limit, four-point function will diverge whenever the exponent satisfies the condtion
\beq
2-\frac{(\vec{p_1}-\vec{p_4})^2}{2} \ge 1 \, \, \Longleftrightarrow  \,\, (\vec{p_1}-\vec{p_4})^2 \le 2.\nonumber
\eeq
\noindent
Now for the kinematic configuration in Case-2, we have three boundaries\footnote{The following discussion assumes $m_1+m_2 < m_3+m_4$. When $m_1+m_2 > m_3+m_4$, essentially the same discussion goes through once we replace $v_{4+}$ with 1.}, which are $v_4=v_{4+}$, $v_4=v_{4B}$, and $v_4= +\infty$. The integral $I_{3}$ exists in the interval $(v_{4+}, +\infty)$, and the integral $I_{1}$ exists in the interval $(v_{4+}, v_{4B})$. Using similar analysis as for Case-1, we find that $I_{3}$ diverges in the same way as for the Case-1 in $v_4 \rightarrow \infty$ limit, that is, whenever $(\vec{p_1}-\vec{p_3})^2 \le 2$. For {\em generic} values of the kinematic parameters in Case-2 configuration, the Mandelstam invariants are finite in the $v_4 \rightarrow v_{4+}$ limit and so both $I_{1}$ and $I_{3}$ do not diverge. Though, $I_{1}$ diverges as $v_4 \rightarrow v_{4B}$ 
whenever $(\vec{p_1}-\vec{p_4})^2 \le 2$.
\\

\subsection{Type-2: Poles at Boundary}
In the previous subsection, we showed that some Mandelstam invariants approach finite values near the boundaries of the integral, for instance, $t \rightarrow -(\vec{p_1}-\vec{p_3})^2$ as $v_4 \rightarrow \infty$, and $u \rightarrow -(\vec{p_1}-\vec{p_4})^2$ as $v_4 \rightarrow v_{4B}$ for the kinematic configuration in Case-1. If corresponding to these finite values, the arguments of gamma functions are non-negative integers, we would have poles at the boundary of the integral. Integration to such poles would not be finite as they are only being approached from one direction in the sense of CPV.

For both Case-1 \& 2 at $v_4=v_{4B}$ we have $u=-(\vec{p_1}-\vec{p_4})^2$ and as $v_4 \rightarrow \infty$ we have $t \rightarrow -(\vec{p_1}-\vec{p_3})^2$, respectively. Corresponding to these values of $t$ and $u$, we would have a boundary pole at $v_4=v_{4B}$ whenever $(\vec{p_1}-\vec{p_4})^2 =0 \ {\rm or}\ 4$, or at $v_4 \rightarrow \infty$  whenever $(\vec{p_1}-\vec{p_3})^2 =0 \ {\rm or}\ 4$ \footnote{This boundary pole for $v_4 \rightarrow \infty$ could also be thought of as a boundary pole at $v_4=0$ if we think of it as a divergence in the integral $I_1$ instead of $I_3$. Such translations about the origin of the divergence exists in other cases as well, but we have chosen not to emphasize them.}. 

One potential divergence that can arise in the scattering amplitude comes from the
\bea
\frac{1}{|v_2^2- v_3^2|} \nonumber
\eea
factor in the integrand at the lower limit of integration, $v_{4+}$. But the leading behavior of this factor goes as $ \sim \frac{1}{(v_4-v_{4+})^{1/2}}$
for {\em generic} $v_{4+}$ and so the integral is finite at the lower limit of integration. 
The rest of the amplitude integrand goes to a constant as $v_{4}\rightarrow v_{4+}$ (except at isolated values of the kinematic parameters which we will discuss momentarily). 
But when $v_{4+}=1$, which corresponds to special choice of parameters, we will see in the next subsection that there is an extra divergence that emerges, which was noted in \cite{Ben}.  

In Case-2, the Mandelstam invariants are finite at $v_4=v_{4+}$ and can be calculated to be
\bea
& & s(v_{4+})\equiv-\frac{m_2 \left(m_1^2+m_2^2+m_3^2-m_4^2-2 m_2 m_3\right)}{m_2-m_3}-2 P_{12}-8 , \label{sboundary} \\
& & t(v_{4+})\equiv-\frac{m_3 \left(m_1^2+m_2^2+m_3^2-m_4^2-2 m_2 m_3\right)}{m_3-m_2}+2 P_{13}-8 , \label{tboundary} \\
& & u(v_{4+})\equiv-m_1^2+m_2^2+m_3^2-m_4^2-2 m_2 m_3+2 P_{14}-8. \label{uboundary}
\eea
\noindent
If these values correspond to poles of the gamma function (i.e., when these expressions equal $4(n-1)$ for some non-negative integer $n$), naively we could expect boundary divergences at $v_4=v_{4+}$. But there is a subtlety involved here. This is because the integral in Case-2 involves two pieces (\ref{case2intsum}) and they contribute destructively at the boundary. One way to see this is to note that the two integrals can locally (around $v_{4+}$) be written as integrals over $v_2$ instead of as integrals over $v_4$, by a change of variables\footnote{A related observation is that the tangent of $s$ (or $v_2$) at $v_{4+}$ is vertical. This means that finding analytic estimates of the integral via Taylor expansion in $v_4$ around $v_{4+}$ is not feasible, even if the integral were perfectly well-defined and finite. So to apply the type of logic that we we used at the beginning of section 3.2 based on the gamma function and its poles, one needs to first go over to a more convenient variable around $v_{4+}$. This is precisely what $v_2$ is.}. The point $v_{4+}$ is a branch point for $v_2$ where the two branches $v_{2u}$ and $v_{2d}$ meet, as can be seen from figure 4. 
\begin{figure}[h]
\centering
\includegraphics[width=0.8\textwidth]{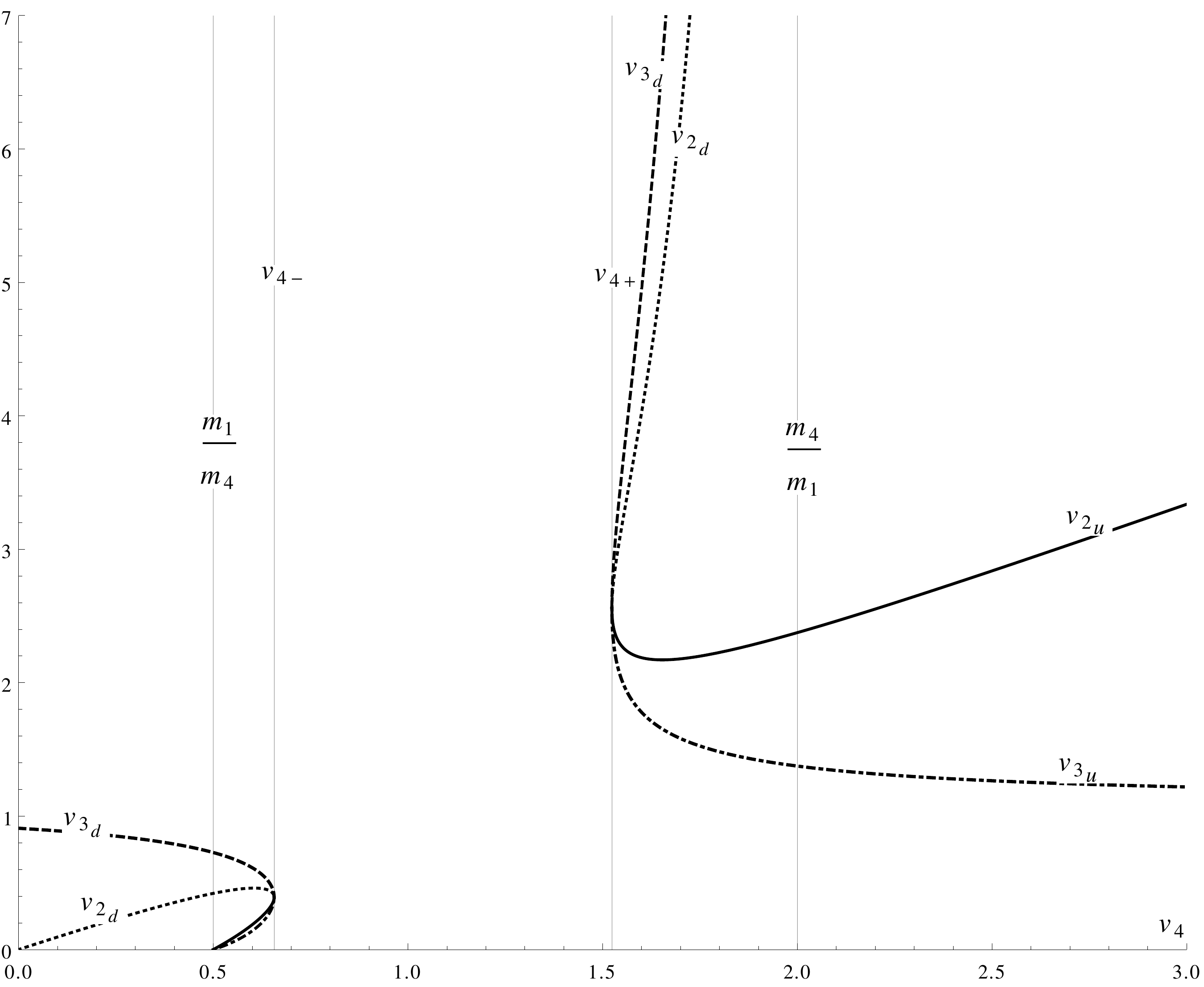}
\caption{The plots of the branches of $v_2$ and $v_3$. Only positive values are relevant, so we plot only them. The values of the kinematic parameters used for making the plot are $m_1=5.0,\ m_2=9.5,\ m_3=5.5,\ m_4=10.0,\ P_{12}=39.5, \ P_{13}=4.0$.}
%\label{fig:circles}
\end{figure}
A basic observation is that the integrands $i_1$ and $i_3$ are the same function, but over the two branches. 

Now consider a situation when there is a pole at the boundary arising from (\ref{sboundary}), i.e., in the gamma function associated to $s$ (or equivalently, $v_2$). The situation when there is a pole in $t$, namely (\ref{tboundary}), is entirely analogous.\footnote{The $u$-channel case (\ref{uboundary}) is different, and we will discuss it towards the end of this subsection.} The potential divergences in $i_1$ and $i_3$ together can be written as
\beq
 \int_{v_{4+}}
 \frac{dv_{4}}{\sqrt{v_4-v_{4+}}} \, \tilde \Gamma (s(v_{2u}(v_4))) +\int_{v_{4+}}
 \frac{dv_{4}}{\sqrt{v_4-v_{4+}}}\, \tilde \Gamma (s(v_{2d}(v_4))) .
 \eeq
We have suppressed the rest of the functions in the integrand because they are well-defined and continuous across the branches and just go along for the ride, and we are using the notation $\tilde\Gamma(s)$ to mean $\Gamma(-1-s/4)$ for brevity. As discussed, $v_2$ is a better variable to work with around $v_{4+}$ so the integral becomes
\bea
\int_{v_{2}(v_{4+})}
\frac{ dv_{2u}}{\sqrt{v_4-v_{4+}}} \,  \frac{\tilde\Gamma (s(v_{2u}))}{dv_{2u}/dv_4} +\int_{v_2(v_{4+})}
\frac{ dv_{2d}}{\sqrt{v_4-v_{4+}}}\, \frac{\tilde\Gamma (s(v_{2d}))}{dv_{2d}/dv_4} .
\eea
Now the crucial observation is that $1/(dv_{2}/dv_4)$ is continuous across the branches, but goes {\em through} zero at $v_{4+}$ and cancels the factor $1/\sqrt{v_4-v_{4+}}$ (up to a minus sign on the $v_{2u}$ branch). The whole expression can be written as an integral along the $v_2(v_4)$ curve stretching across both branches, the portion near $v_{4+}$ being proportional to
\bea
 \int dv_{2} \,  \tilde\Gamma (s(v_{2})),
\eea
where now the integral is in a small neighborhood around the point $v_2(v_{4+})$. There is a pole at that point, but this expression is bound to be finite in the sense of CPV by the discussions of the previous sections. That $s$ as a function of $v_2$ does not have a maximum/minimum at $v_2(v_{4+})$ is readily checked\footnote{An exception to this is when we have a maximum/minimum of $s$ or $t$ at $v_{4+}$. This corresponds to the condition $(m_3+m_4)=(m_1+m_2)$ which coincides with $v_{4+}=1$ and gives rise to divergences. In such a situation, the expression (\ref{sboundary}) for gamma function poles for a non-negative integer argument $n$ happens when
\bea
s=(m_1 + m_2)^2 - (\vec p_1 + \vec p_2)^2 = 4 (n-1). \label{spoles}
\eea
Similar arguments apply for $t$ %and we have divergences from gamma function poles at $v_{4+}=1$ for $(m_3+m_4)=(m_1+m_2)$,  
which reduces (\ref{tboundary}) to
\bea
t=(m_1 - m_3)^2 - (\vec p_1 - \vec p_3)^2 = 4 (n-1). \label{tpoles}
\eea
Satisfyingly, both these expressions can be understood as specical cases (for $s$ and $t$) of the extrema poles we discuss in the next subsection  when the extremum happens at the boundary.
%for a non-negative integer $n$.
}. The conclusion of all this is that the potential divergences that could have arisen from (\ref{sboundary}) and (\ref{tboundary}) when they correspond to gamma function poles are actually spurious, and in fact there are no divergences from them (unless the Mandelstam invariants have a maximum/minimum at $v_{4+}$ as discussed in the footnote). 

The conclusion of all this is that the potential divergences that could have arisen from (\ref{sboundary}) and (\ref{tboundary}) when they correspond to gamma function poles are actually spurious, and in fact there are no divergences from them. 

The same is not however true about the poles corresponding to (\ref{uboundary}). One can check that there are indeed divergences in the integral that arise when (\ref{uboundary}) is equal to $4(n-1)$, with $n$ a non-negative integer. This condition can in fact be written in a suggestive form as
\bea
(m_2 - m_3)^2 - (\vec p_2 - \vec p_3)^2 = 4 (n-1). \label{upoles}
\eea

This suggests that this divergence can be understood in terms of intermediate tower of string states going on-shell and is therefore a ``good" divergence unrelated to the singularity.\footnote{It is possible to show that (\ref{upoles}) has solutions in kinematic parameters that satisfy all the constraints for arbitrarily large $n$. An explicit way to see this is to choose an $\alpha$  (where $0 < \alpha < 1/\sqrt{2}$) that solves 
\bea
n =  \frac{(1-8\alpha^2+2\alpha^3+4\alpha^4+4\alpha^5+\alpha^6)}{4 \alpha^2}
\eea
and set
\bea
m_1 = \alpha, \,\, m_4= \frac{1}{\alpha}, \ m_2 = \frac{1}{\alpha}-\alpha, \,\, m_3 = \alpha(1+\alpha), \ P_{12} = \frac{1}{2 \alpha^2 }, \,\, P_{13} = \frac{\alpha^2}{2} +4 
\eea  
It is easily checked that all the consistency conditions between the kinematic parameters are satisfied.}
To argue further for the infrared character of these divergences, one can replace the vertex operators (\ref{vertex}) by their asymptotic behavior for large $X^+X^-$ (far away from the singularity), given above (A.6) of \cite{Ben}, and act with $\Box$ on $\psi_3^*\psi_2$. The result for large $X^+X^-$ is that $\Box=-(m_2 - m_3)^2 + (\vec p_2 - \vec p_3)^2$, in agreement with (\ref{upoles}).

A thing to note about these boundary pole divergences is that the conditions above for them to arise are constraints on the kinematic parameters. This means that they are not generic. 

\subsection{Type-3: Poles at Mandelstam Maxima/Minima}

As we discussed earlier, whenever there is a gamma function pole in $v_4$ space at a maximum/minimum of a Mandelstam invariant, the integration through the pole would cause a divergence in the four-point function.

The discussion of the extrema splits into two possibilities depending on whether $(m_3+m_4)>(m_1+m_2)$ or $(m_3+m_4)<(m_1+m_2)$. For extrema of $s$ with  $(m_3+m_4)>(m_1+m_2)$, the divergences appear from poles in the integration range when $dv_{2}/dv_{4}=0$. In this situation, the expression $-1-\frac{s}{4}=-n$ for non-negative integer $n$, reduces to the suggestive form 
\bea
s=(m_4 + m_3)^2 - (\vec p_4 + \vec p_3)^2 = 4 (n-1). \label{SPOLE1}
\eea
On the other hand when $(m_3+m_4)<(m_1+m_2)$, the relevant maxima/minima of $s$ arise from $ds/dv_{2}=0$ (i.e. $v_2=1$), which again reduces %for gamma function poles 
to a simple form
\bea
s=(m_1 + m_2)^2 - (\vec p_1 + \vec p_2)^2 = 4 (n-1) \label{SPOLE2}
\eea
where $n$ is some non-negative integer. It should be noted that the two conditions above are same whenever $(m_3+m_4)=(m_1+m_2)$, occurring at $v_4=1$.

A similar conclusion holds for $t$. %we find that relavant extrema only exist in Case-2 kinematic configuration. 
Here also, condition $(m_3+m_4)>(m_1+m_2)$ means poles come from $dv_3/dv_4=0$. The fact that these extrema are at non-negative integers $n$ gives rise to
\bea
t=(m_2 - m_4)^2 - (\vec p_2 - \vec p_4)^2 = 4 (n-1). \label{tpoles1}
\eea
An extra caveat is that these poles lie in the integration range only when $m_2 >m_4$ (which automatically forces $m_3 > m_1$).
For $(m_3+m_4)<(m_1+m_2)$, we have extrema of $t$ when $dt/dv_{3}=0$ (i.e. $v_3=1$) giving us the condition on $t$,   
\bea
t=(m_1 - m_3)^2 - (\vec p_1 - \vec p_3)^2 = 4 (n-1). \label{tpoles2}
\eea
These $t$-poles are in the integration range only when $m_3 >m_1$ (which automatically forces $m_2 > m_4)$.
Again, both the former conditions are equivalent for $(m_3+m_4)=(m_1+m_2)$. 

It is worth remarking that the extrema of $u$ occur only at $v_4=1$, and this point is part of the integration range when $(m_3+m_4)\le(m_1+m_2)$. Poles at these extrema will occur whenever $u$ at $v_4=1$ satisfies 
\bea
u=(m_1 - m_4)^2 - (\vec p_1 - \vec p_4)^2 = 4 (n-1) \label{upoles1}
\eea
for non-negative integer $n$. For convenience of comparison of the various towers of divergent poles that we have uncovered, we repeat also (\ref{upoles}) here:
\bea
u=(m_2 - m_3)^2 - (\vec p_2 - \vec p_3)^2 = 4 (n-1). \label{upoles2}
\eea
Note that these poles are not extrema poles, and exist only when $(m_3+m_4)\ge(m_1+m_2)$ which is the necessary condition for boundary poles to exist. 

Now we see that we have towers of divergent poles for each channel. %, we have divergences which are fully symmetric in their labels. 
The arguments for the existence of the various poles is not manifestly symmetric under interchange of labels because we chose to do our integration along $v_4$, which breaks the symmetry between $v_3$ and $v_4$, and also because we are looking at various cases separately (in particular, note that the cases we are explicitly considering have $m_4 \ge m_3$ and exchanging $t$ and $u$ corresponds to exchanging $m_4$ and $m_3$). The origin of a divergence as a boundary/extremum divergence is not independent of these choices and is merely an artifact. Another sanity check is that when the parameter condition $(m_3+m_4)=(m_1+m_2)$ is satisfied, it is straightforward to see that the extrema divergences become the boundary divergences of the last section 
 %(\ref{SPOLE1}-\ref{upoles1}) are all precisely the same as (\ref{spoles}-\ref{upoles}), respectively. \
because $v_{4+}=1$ in this case.%, and so the boundary divergences are also extremum type divergences.

%But the other divergences we find arising from (\ref{SPOLE1}-\ref{SPOLE2}) and (\ref{tpoles1}-\ref{tpoles2}) are new. 
These divergences that we have identified are expected IR divergences corresponding to intermediate string states going on-shell. We note that these divergences are non-generic (as in, they appear only when the paramters satisfy the conditions listed in this subsection). But it is possible to show that solutions exist for all $n$ (non-negative). This is somewhat non-trivial to demonstrate in general because the kinematic paramters have to satisfy various constraints as well as consistency conditions between them. For the case of the $s$-poles (\ref{SPOLE1}), for example, %it is straightforward to see that there are infinitely many solutions (\ref{SPOLE1}) for arbitrarily large $n$: 
once one finds a solution (by trial and error - this can be easily accomplished with Mathematica) for a small integer $n$ that satisfies the constraints, one can increase $m_4$ as one wishes while holding the other parameters fixed in order to satisfy (\ref{SPOLE1}): it is easy to check that all the kinematic constraints will still be satisfied. Various arguments of a similar flavor can also be used for (\ref{SPOLE2}-\ref{upoles2}) as well\footnote{A specific ansatz for $u$-poles was given in a previous footnote. The $t$-poles (\ref{tpoles1}) are also somewhat subtle, so we present an explicit ansatz here that finds solutions of (\ref{tpoles1}) for all non-negative integers. %It is possible to show that (\ref{tpoles1}) has solutions in kinematic parameters that satisfy all the constraints for arbitrarily large $n$. 
Choose an $\epsilon$  (where $0 < \epsilon \le \frac{1}{4}$) that solves 
\bea
n = -1 +  \frac{1}{16 \epsilon^2}
\eea
and set
\bea
m_1 = \epsilon, \,\, m_4= \frac{1}{\epsilon^{2}}, \ m_2 = \frac{1}{\epsilon^{2}}+\frac{1}{2\epsilon}, \,\, m_3 = \frac{1}{\epsilon}, \ P_{12} = \frac{1}{2 \epsilon^{2} }, \,\, P_{13} = \frac{\epsilon^2}{2} +\frac{1}{2 \epsilon^{2}} 
\eea  
It is easily checked that all the consistency conditions for the kinematic parameters are satisfied.} to show that these expressions correspond to infinite towers of string states going on-shell.

%Only $s$ and $u$ have relevant extrema, so one can show that this type of divergence of the four-point function will occur when either \beq n=\frac{1}{4} (-4 -m_1^2 -m_2^2 +m_3^2 +m_4^2 +2m_3 m_4- 2P_{12})  \label{spoles} \eeq  or  \beq n=\frac{1}{4}(-4-2m_1m_4+2P_{14}),  \label{upoles1} \eeq where $n$ is some non-negative integer. It is worth remarking that the extrema of $u$ occur at $v_4=1$, and this point is part of the integration range when $(m_4-m_1)=(m_2-m_3)$ or when $m_4 =m_1$. When the former condition is satisfied, one can show that the condition (\ref{upoles}) is precisely the same as (\ref{upoles1}). This is because $v_{4+}=1$ in this case, so the boundary divergence is also an extremum divergence so these are basically a special case of the divergences we saw in (\ref{upoles}).  But the divergences arising from (\ref{spoles}) are new divergences. Again remarkably, we can re-write this condition in the suggestive form \bea (m_4 + m_3)^2 - (\vec p_4 + \vec p_3)^2 = 4 (n-1), \label{SPOLE} \eea implying again that these are expected divergences corresponding to string states going on-shell. It is straightforward to construct solutions of this equation in kinematic parameters for arbitrarily large $n$: once one finds a solution for a small $n$ by trial and error, one can increase $m_4$ as one wishes while holding the other parameters fixed in order to satisfy (\ref{spoles}) and all the constraints will be satisfied. Again, we note that these divergences are non-generic, even though there are a countably infinite number of them in the kinematic parameter space.

\subsection{Type-4: IR divergences Identified in \cite{Ben}}

For the Case-2 configuration, we find that $v_{4+}=v_{4-}=1$ whenever $(m_1+m_2)=(m_3+m_4)$ and this gives rise to one more divergence\footnote{This divergence was noticed in \cite{Ben}. See their discussion right before section 4. The various sign choices arising in their $(m_4-m_1)^2=(m_2-m_3)^2$ condition are distributed over the various cases here, which are analogous to this one. So we do not discuss them.}. 
As $v_4 \rightarrow 1$, both $v_{2u}, v_{3u} \rightarrow 1$ and all the Mandelstam invariants are finite. The four-point function integral takes the form
\beq
C^{''} \times \int_{v_4=1}^{} dv_4 \, \frac{1}{(v_{2u}-v_{3u})} , 
\eeq
where  $C^{''}$ is a constant. This integral above diverges logarithmically in the $v_4 \rightarrow 1$ limit. 
 This divergence is not associated with the singularity but is an infra-red divergence \cite{Ben}. 

It is important to note that the condition $v_{4+}=1$ is crucial for the existence of this divergence: $v_{4+}=1$ forces $v_{4+}=v_{4-}$ and this results in $\Delta$ in the definition of $v_2$ and $v_3$ contributing to the leading behavior. This makes the integrand near the lower limit to behave as $1/(v_4-v_{4+})$ instead of $1/(v_4-v_{4+})^{1/2}$. Crucially, this means that the divergence again has an interpretation as an isolated IR divergence \cite{Ben}. 

An integral can diverge only from one of its boundaries or from a region in the integration range where the integrand blows up badly enough. We have checked every such possibility by looking at the boundaries, the poles of the gamma functions and the divergence due to $1/(v_{2u,d}-v_{3u,d})$. This means that our scan of the divergences is an exhaustive one.

\section{Summary of Divergences}

In this section, we summarize the divergences in the tree-level 2-to-2 string scattering amplitude. We will only list the divergences in the context of the two cases we have considered. There exist analogous divergences that arise when the ordering of the mass parameters are changed, but these can all be obtained from the cases we list here by appropriate permutations of the labels. 
\begin{itemize}
\item UV divergences that arise when $(\vec p_1 - \vec p_3)^2 \le 2$  
or when  $(\vec p_1 - \vec p_4)^2 \le 2$.\footnote{When one re-instates the $\alpha'$ these will read $\alpha' (\vec p_1 - \vec p_3)^2 \le 2$ and $\alpha' (\vec p_1 - \vec p_4)^2 \le 2$.} 
\item  IR divergences that arise when one of the conditions (\ref{SPOLE1}-\ref{upoles2}) is satisfied, arising from the tower of string states going on-shell.
\item  IR divergences arising from tachyons and massless states going on-shell when $(\vec p_1 - \vec p_3)^2 =0,\ 4$ or when $(\vec p_1 - \vec p_4)^2 =0,\  4$.
\item The logarithmic IR divergence that arises when $(m_1+m_2)=(m_3+m_4)$.
\end{itemize}

The first and last kinds of divergences were already noticed in \cite{Ben}. Our claim here is that all the other IR divergences are physical divergences expected to be present in healthy scattering amplitudes. 
The UV divergences disappear in the large (dimensionless) $\alpha'$ region of the parameter space, as already belabored in the introduction.

\acknowledgments
We thank Pallab Basu for managing to feed one of the authors for the last few months, despite Indian bureaucracy. CK thanks Joan Simon for discussions on a related collaboration. This work was supported in part by the Belgian Federal Science Policy Office through the Interuniversity Attraction Pole P7/37, by FWO-Vlaanderen through project G020714N,  and by the Vrije Universiteit Brussel through the Strategic Research Program ``High-Energy Physics''.\\

\bibliographystyle{JHEP}
\bibliography{AyushBen}

\providecommand{\href}[2]{#2}\begingroup\raggedright\begin{thebibliography}{10}

\bibitem{Ben}
M.~Berkooz, B.~Craps, D.~Kutasov, and G.~Rajesh, {\it {Comments on cosmological
  singularities in string theory}},  {\em JHEP} {\bf 0303} (2003) 031,
  [\href{http://xxx.lanl.gov/abs/hep-th/0212215}{{\tt hep-th/0212215}}].

\bibitem{Liu:2002yd}
H.~Liu, G.~W. Moore, and N.~Seiberg, {\it {The Challenging cosmic
  singularity}},  \href{http://xxx.lanl.gov/abs/gr-qc/0301001}{{\tt
  gr-qc/0301001}}.

\bibitem{Cornalba:2003kd}
L.~Cornalba and M.~S. Costa, {\it {Time dependent orbifolds and string
  cosmology}},  {\em Fortsch.Phys.} {\bf 52} (2004) 145--199,
  [\href{http://xxx.lanl.gov/abs/hep-th/0310099}{{\tt hep-th/0310099}}].

\bibitem{Durin:2005ix}
B.~Durin and B.~Pioline, {\it {Closed strings in Misner space: A Toy model for
  a big bounce?}},  \href{http://xxx.lanl.gov/abs/hep-th/0501145}{{\tt
  hep-th/0501145}}.

\bibitem{Craps:2006yb}
B.~Craps, {\it {Big Bang Models in String Theory}},  {\em Class.Quant.Grav.}
  {\bf 23} (2006) S849--S881,
  [\href{http://xxx.lanl.gov/abs/hep-th/0605199}{{\tt hep-th/0605199}}].

\bibitem{Berkooz:2007nm}
M.~Berkooz and D.~Reichmann, {\it {A Short Review of Time Dependent Solutions
  and Space-like Singularities in String Theory}},  {\em Nucl.Phys.Proc.Suppl.}
  {\bf 171} (2007) 69--87, [\href{http://xxx.lanl.gov/abs/0705.2146}{{\tt
  arXiv:0705.2146}}].

\bibitem{LMS}
H.~Liu, G.~W. Moore, and N.~Seiberg, {\it {Strings in a time dependent
  orbifold}},  {\em JHEP} {\bf 0206} (2002) 045,
  [\href{http://xxx.lanl.gov/abs/hep-th/0204168}{{\tt hep-th/0204168}}].

\bibitem{Lawrence:2002aj}
A.~Lawrence, {\it {On the Instability of 3-D null singularities}},  {\em JHEP}
  {\bf 0211} (2002) 019, [\href{http://xxx.lanl.gov/abs/hep-th/0205288}{{\tt
  hep-th/0205288}}].

\bibitem{HoroPolch}
G.~T. Horowitz and J.~Polchinski, {\it {Instability of space - like and null
  orbifold singularities}},  {\em Phys.Rev.} {\bf D66} (2002) 103512,
  [\href{http://xxx.lanl.gov/abs/hep-th/0206228}{{\tt hep-th/0206228}}].

\bibitem{Shubho1}
C.~Krishnan and S.~Roy, {\it {Higher Spin Resolution of a Toy Big Bang}},  {\em
  Phys.Rev.} {\bf D88} (2013) 044049,
  [\href{http://xxx.lanl.gov/abs/1305.1277}{{\tt arXiv:1305.1277}}].

\bibitem{Shubho2}
C.~Krishnan and S.~Roy, {\it {Desingularization of the Milne Universe}},
  \href{http://xxx.lanl.gov/abs/1311.7315}{{\tt arXiv:1311.7315}}.

\bibitem{Rajesh}
M.~R. Gaberdiel and R.~Gopakumar, {\it {An AdS$_3$ Dual for Minimal Model
  CFTs}},  {\em Phys.Rev.} {\bf D83} (2011) 066007,
  [\href{http://xxx.lanl.gov/abs/1011.2986}{{\tt arXiv:1011.2986}}].

\bibitem{Campoleoni}
A.~Campoleoni, S.~Fredenhagen, S.~Pfenninger, and S.~Theisen, {\it {Asymptotic
  symmetries of three-dimensional gravity coupled to higher-spin fields}},
  {\em JHEP} {\bf 1011} (2010) 007,
  [\href{http://xxx.lanl.gov/abs/1008.4744}{{\tt arXiv:1008.4744}}].

\bibitem{Avinash}
C.~Krishnan, A.~Raju, and S.~Roy, {\it {A Grassmann path from $AdS_3$ to flat
  space}},  {\em JHEP} {\bf 1403} (2014) 036,
  [\href{http://xxx.lanl.gov/abs/1312.2941}{{\tt arXiv:1312.2941}}].

\bibitem{Arjun}
H.~Afshar, A.~Bagchi, R.~Fareghbal, D.~Grumiller, and J.~Rosseel, {\it {Spin-3
  Gravity in Three-Dimensional Flat Space}},  {\em Phys.Rev.Lett.} {\bf 111}
  (2013), no.~12 121603, [\href{http://xxx.lanl.gov/abs/1307.4768}{{\tt
  arXiv:1307.4768}}].

\bibitem{Troncoso}
H.~A. Gonzalez, J.~Matulich, M.~Pino, and R.~Troncoso, {\it {Asymptotically
  flat spacetimes in three-dimensional higher spin gravity}},  {\em JHEP} {\bf
  1309} (2013) 016, [\href{http://xxx.lanl.gov/abs/1307.5651}{{\tt
  arXiv:1307.5651}}].

\bibitem{Vasiliev}
X.~Bekaert, S.~Cnockaert, C.~Iazeolla, and M.~Vasiliev, {\it {Nonlinear higher
  spin theories in various dimensions}},
  \href{http://xxx.lanl.gov/abs/hep-th/0503128}{{\tt hep-th/0503128}}.

\bibitem{Sundborg}
B.~Sundborg, {\it {Stringy gravity, interacting tensionless strings and
  massless higher spins}},  {\em Nucl.Phys.Proc.Suppl.} {\bf 102} (2001)
  113--119, [\href{http://xxx.lanl.gov/abs/hep-th/0103247}{{\tt
  hep-th/0103247}}].

\bibitem{Witten}
E.~Witten, {\it {Talk given at J.H. Schwarz 60th Birthday Conference}}, .

\bibitem{Giombi:2011kc}
S.~Giombi, S.~Minwalla, S.~Prakash, S.~P. Trivedi, S.~R. Wadia, et~al., {\it
  {Chern-Simons Theory with Vector Fermion Matter}},  {\em Eur.Phys.J.} {\bf
  C72} (2012) 2112, [\href{http://xxx.lanl.gov/abs/1110.4386}{{\tt
  arXiv:1110.4386}}].

\bibitem{Chang:2012kt}
C.-M. Chang, S.~Minwalla, T.~Sharma, and X.~Yin, {\it {ABJ Triality: from
  Higher Spin Fields to Strings}},  {\em J.Phys.} {\bf A46} (2013) 214009,
  [\href{http://xxx.lanl.gov/abs/1207.4485}{{\tt arXiv:1207.4485}}].

\bibitem{Gross:1987kza}
D.~J. Gross and P.~F. Mende, {\it {The High-Energy Behavior of String
  Scattering Amplitudes}},  {\em Phys.Lett.} {\bf B197} (1987) 129.

\end{thebibliography}\endgroup

\end{document}